\documentclass{PoS}

\usepackage{graphicx}
\usepackage{amssymb}
\usepackage{amsmath}

\newcommand{\beq}{\begin{eqnarray}}
\newcommand{\eeq}{\end{eqnarray}}

\title{(2+1)-flavor QCD Thermodynamics from the Gradient Flow
\thanks{Numerical simulation for this study was carried out on Hitachi SR16000 
and IBM System Blue Gene Solution at KEK under its Large-Scale Simulation Program
(No. 14/15-11), Hitachi SR16000 at YITP Kyoto University and NEC SX-8 and SX-9 at RCNP Osaka University.  
E.~I. is supported in part 
by Strategic Programs for Innovative Research
(SPIRE) Field 5.  
H.~S. and T.~U. are in supported by the Japan Society for the Promotion of Science (JSPS) Grants Number 23540330 and 26400251, respectively.
}}

\ShortTitle{(2+1)-flavor QCD Thermodynamics from the Gradient Flow}

\author{\speaker{Etsuko Itou}\\
         Theory Center, High Energy Accelerator Research Organization (KEK) \\
        E-mail: \email{eitou@post.kek.jp}}

\author{Hiroshi Suzuki\\
        Department of Physics, Kyushu University, 744 Motooka, Nishi-ku, Fukuoka, 812-0395, Japan\\
        E-mail: \email{hsuzuki@phys.kyushu-u.ac.jp}}

\author{Yusuke Taniguchi\\
        Graduate School of Pure and Applied Sciences, University of Tsukuba, Tsukuba, Ibaraki 305-8571, Japan\\
        E-mail: \email{tanigchi@het.ph.tsukuba.ac.jp }}

\author{Takashi Umeda\\
        Graduate School of Education, Hiroshima University, Hiroshima 739-8524, Japan\\
        E-mail: \email{tumeda@hiroshima-u.ac.jp}}

\abstract{
Recently, we proposed a novel method to define and calculate the energy-momentum tensor (EMT) in lattice gauge theory on the basis of the Yang-Mills gradient flow~\cite{Asakawa:2013laa}.
In this proceedings, we summarize the basic idea and technical steps to obtain the bulk thermodynamic quantities in lattice gauge theory using this method for the quenched and $(2+1)$-flavor QCD.
The revised results of integration measure (trace anomaly) and entropy density of the quenched QCD with corrected coefficients are shown.
Furthermore, we also show the flow time dependence of the parts of EMT including the dynamical fermions.
This work is based on a joint-collaboration between FlowQCD and WHOT QCD.
}

\FullConference{The 33rd International Symposium on Lattice Field Theory\\
		14 -18 July 2015\\
		Kobe International Conference Center, Kobe, Japan*}

\begin{document}

\section{Introduction}
Measurement of the energy-momentum tensor (EMT) that is a generator of the general coordinate transformation is difficult using the lattice numerical simulation, since the lattice regularization manifestly breaks the corresponding invariance.
Recently, we proposed a novel method to obtain the EMT using the lattice numerical simulation~\cite{Asakawa:2013laa} based on the small flow-time expansion  of the Yang-Mills gradient flow~\cite{Luscher:2010iy, Suzuki:2013gza}.
As a first observation, the bulk thermal quantities, namely integration measure (trace anomaly) and thermal entropy density, are calculated by the direct measurement of EMT.
In this proceedings, we review the quenched QCD results, and also show the detailed strategy and a preliminary result for $(2+1)$-flavor QCD simulation.

\section{Quenched QCD}
\subsection{strategy}
A key property of the Yang-Mills gradient flow~\cite{Luscher:2010iy} is UV finiteness of local operators~\cite{Luscher:2011bx}.
For example, the following
gauge-invariant local products of dimension~$4$ are UV finite for the positive flow-time ($t>0$):
$U_{\mu\nu}(t,x)\equiv G_{\mu\rho}(t,x)G_{\nu\rho}(t,x)
-\frac{1}{4}\delta_{\mu\nu}G_{\rho\sigma}(t,x)G_{\rho\sigma}(t,x)$
and~$E(t,x)\equiv\frac{1}{4}G_{\mu\nu}(t,x)G_{\mu\nu}(t,x)$.
Here $G_{\mu \nu}$ presents the field strength constructed by the flowed gauge field.

The expansion coefficients are governed by
the renormalization group equation and their small $t$ behavior can be
calculated by perturbation theory thanks to the asymptotic freedom. For the
operators mentioned above, we have~\cite{Suzuki:2013gza,DelDebbio:2013zaa}
\begin{align}
   U_{\mu\nu}(t,x)
   &=\alpha_U(t)\left[
   T_{\mu\nu}^R(x)-\frac{1}{4}\delta_{\mu\nu}T_{\rho\rho}^R(x)\right]
   +O(t),
\label{eq:(2)}\\
   E(t,x)
   &=\left\langle E(t,x)\right\rangle_0
   +\alpha_E(t)T_{\rho\rho}^R(x)
   +O(t),
\label{eq:(3)}
\end{align}
where $\langle\cdot\rangle_0$ is vacuum expectation value (v.e.v.) and $T_{\mu\nu}^R(x)$
is the correctly-normalized conserved EMT with its vacuum expectation value
subtracted. Abbreviated are the contributions from the operators of
dimension~$6$ or higher, which are suppressed for small~$t$.

In $t \rightarrow 0$ limit, the correctly-normalized EMT is given by
\begin{align}
   T_{\mu\nu}^R(x)
   =\lim_{t\to0}\left\{\frac{1}{\alpha_U(t)}U_{\mu\nu}(t,x)
   +\frac{\delta_{\mu\nu}}{4\alpha_E(t)}
   \left[E(t,x)-\left\langle E(t,x)\right\rangle_0 \right]\right\},
\label{eq:(4)}
\end{align}
where the perturbative coefficients are found to be~\cite{Suzuki:2013gza}
\begin{align}
   \alpha_U(t)
   &=\Bar{g}(1/\sqrt{8t})^2
   \left[1+2b_0\Bar{s}_1\Bar{g}(1/\sqrt{8t})^2+O(\Bar{g}^4)\right],
\label{eq:(5)}
\\
   \alpha_E(t)
   &=\frac{1}{2b_0}\left[1+2b_0\Bar{s}_2
   \Bar{g}(1/\sqrt{8t})^2+O(\Bar{g}^4)\right].
\label{eq:(6)}
\end{align}
Here $\bar{g}(q)$, which is the running gauge coupling constant, and the coefficients $\Bar{s}_1$ and $\Bar{s}_2$ depend on the renormalization scheme.
In the $\overline{\text{MS}}$ scheme with the scale $q=1/\sqrt{8t}$,
$\Bar{s}_1=\frac{7}{22}+\frac{1}{2}\gamma_E-\ln2\simeq -0.08635752993$,
$\bar{s}_2=\frac{21}{44}-\frac{b_1}{2b_0^2}=\frac{27}{484}\simeq0.05578512397$,\footnote{Note that in the published version of our paper~\cite{Asakawa:2013laa} the values of these coefficients were wrong. Corrected results are shown in Ver.3 on arXiv.}
with 
$b_0=\frac{1}{(4\pi)^2}\frac{11}{3}N_c$,
$b_1=\frac{1}{(4\pi)^4}\frac{34}{3}N_c^2$,
and~$N_c=3$. 

Our procedure to calculate the EMT on the lattice has the
following four steps:\\
\textbf{Step 1:} Generate gauge configurations at~$t=0$ on a space-time lattice
with the lattice spacing~$a$ and the lattice size~$N_s^3\times N_\tau$.\\
\textbf{Step 2:} Solve the gradient flow for each configuration to obtain the
flowed link variables in the fiducial window, $a\ll\sqrt{8t}\ll R$. Here,
$R$~is an infrared cutoff scale such as~$\Lambda_{\text{QCD}}^{-1}$
or~$T^{-1}=N_\tau a$. The first (second) inequality is necessary to suppress 
finite $a$ corrections (non-perturbative corrections and finite volume
corrections).\\
\textbf{Step 3:} Construct $U_{\mu\nu}(t,x)$ and~$E(t,x)$
in~Eqs.~(\ref{eq:(2)}) and~(\ref{eq:(3)}) in terms of the flowed link variables
and average over the gauge configurations at each~$t$.\\
\textbf{Step 4:} 
Carry out an extrapolation toward $(a,t)=(0,0)$, first $a\to0$ and then $t\to0$ under the condition in \textbf{Step 2}.

\subsection{results}
We consider the pure
$SU(3)$ gauge theory defined on a four-dimensional Euclidean lattice, whose
thermodynamics has been extensively studied by the integral
method~\cite{Boyd:1996bx,Okamoto:1999hi,Umeda:2008bd,Borsanyi:2012ve}. 
We utilize the Wilson plaquette gauge action under the periodic
boundary condition on~$N_s^3\times N_\tau=32^3\times(6,8,10)$ lattices with
several different $\beta=6/g_0^2$ ($g_0$~being the bare coupling constant) shown in Table~\ref{table:Nt_beta}.
These lattice parameters are determined by the relation between the Sommer scale and $\beta$ given by ALPHA collaboration~\cite{Guagnelli:1998ud}
and the critical temperature ($T_c$) given in Ref.~\cite{Boyd:1996bx}.
Gauge configurations are generated by the pseudo-heatbath algorithm with the
over-relaxation, mixed in the ratio of~$1:5$. We call one pseudo-heatbath
update sweep plus five over-relaxation sweeps as a ``Sweep''. To eliminate the
autocorrelation, we take $200$--$500$ Sweeps between measurements. The number
of gauge configurations for the measurements at finite~$T$ is~$100$--$300$.
Statistical errors are estimated by the jackknife method.

\begin{table}[h]
\begin{center}
\begin{tabular}{|c||c|c|c||c|}
\hline
$N_\tau$ & 6 & 8 & 10 & $T/T_c$ \\
\hline
         & 6.20 & 6.40 & 6.56 &  1.65 \\ 
$\beta$  & 6.02 & 6.20 & 6.36 &  1.24 \\
         & 5.89 & 6.06 & 6.20 &  0.99 \\
\hline
\end{tabular}
\caption{Values of $\beta$ and~$N_\tau$ for each temperature.}
\label{table:Nt_beta}
\end{center}
\end{table}

\begin{figure}[t]
\begin{center}
\includegraphics[scale=0.45]{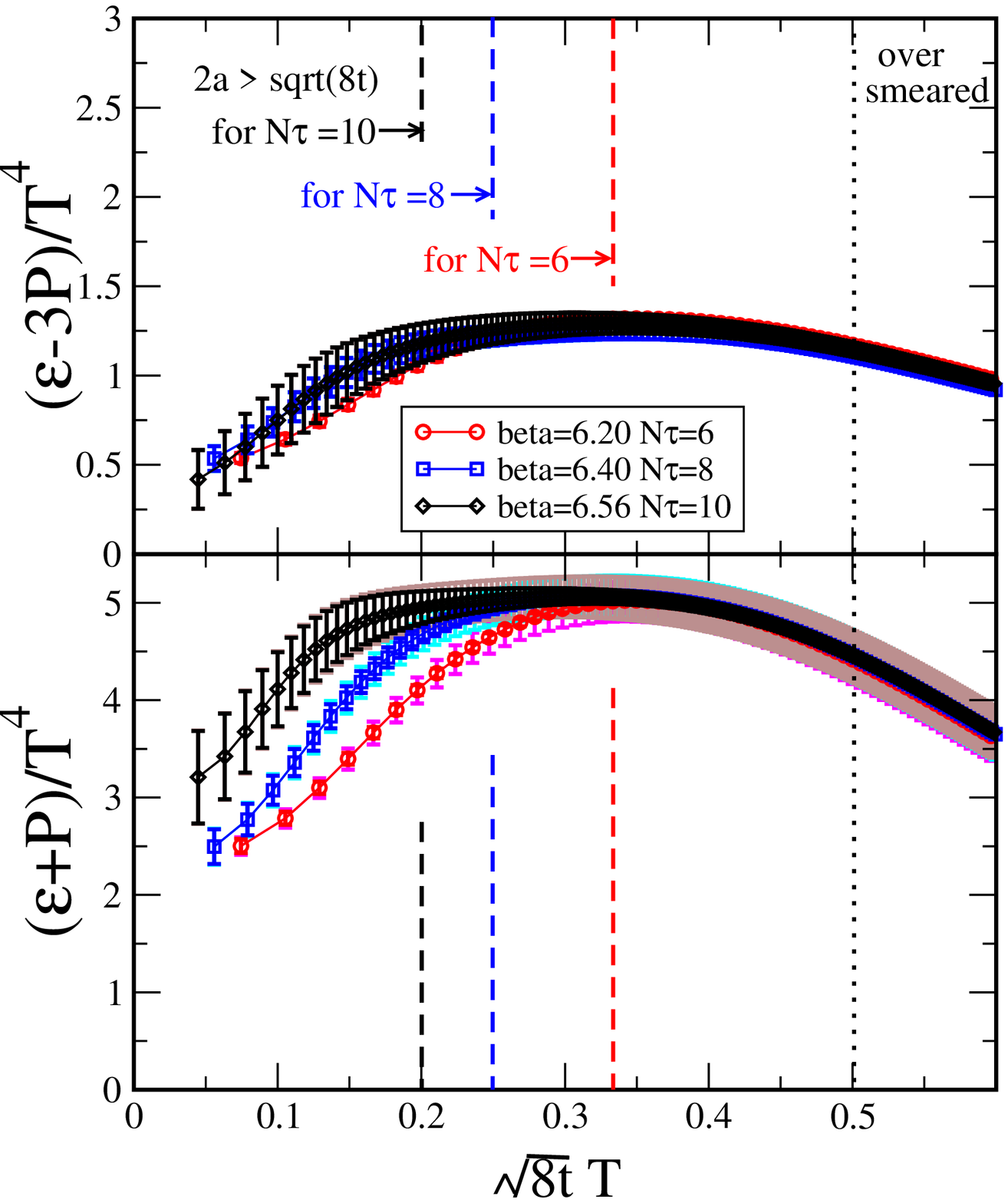}
\includegraphics[scale=0.45]{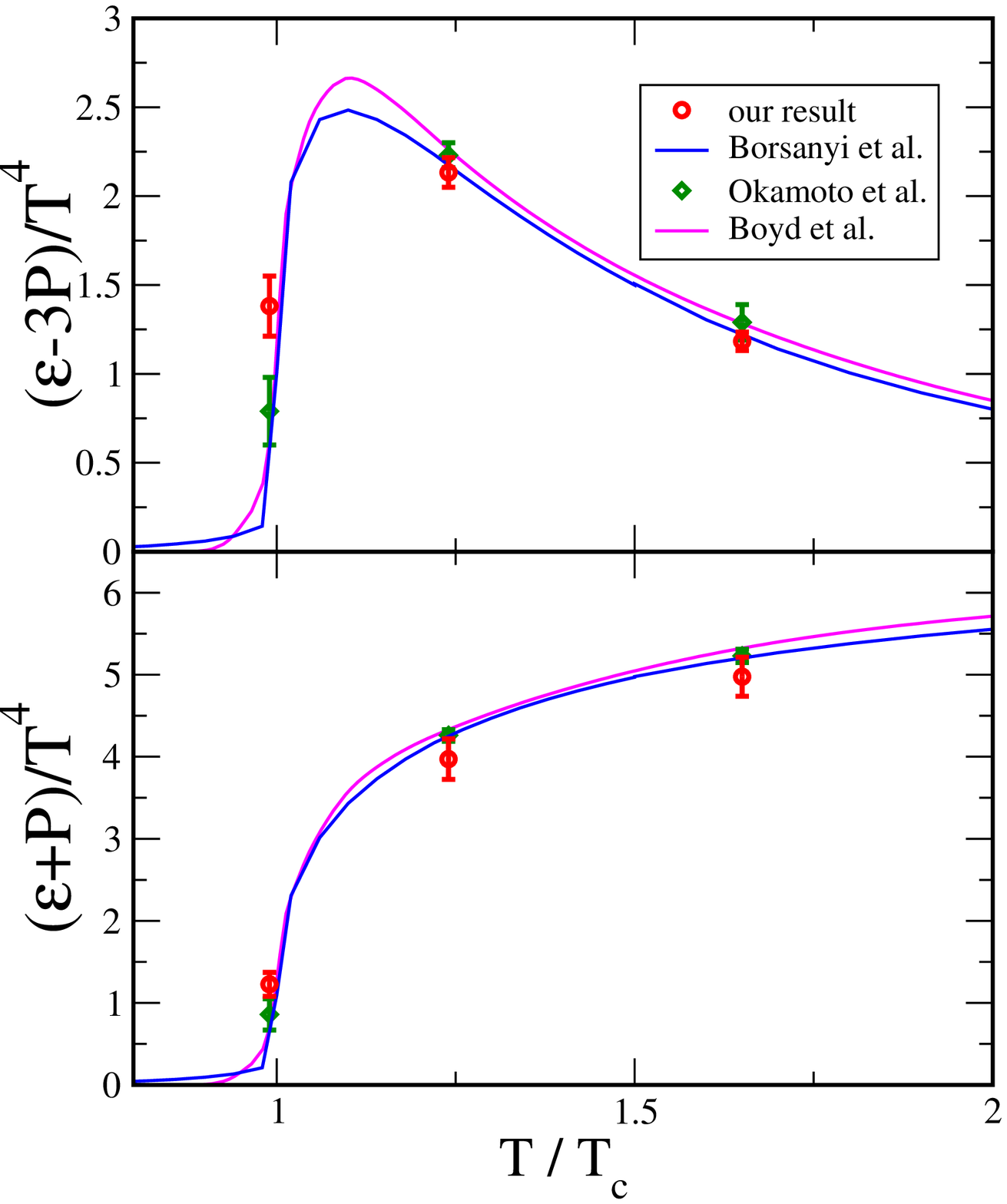}
\caption{(left)Flow time dependence of the dimensionless interaction measure (top
panel) and the dimensionless entropy density (bottom panel) for different
lattice spacings at fixed~$T/T_c=1.65$. The circles (red) the squares (blue),
and the diamonds (black) correspond to~$N_\tau=6$, $8$, and~$10$, respectively.
The bold error bars denote the statistical errors, while the thin error bars
(brown, cyan, and magenta) include both statistical and systematic errors.
(right)Continuum limit of the interaction measure and entropy density
obtained by the gradient flow for~$T/T_c=1.65$, $1.24$, and~$0.99$ with $300$
gauge configurations. The magenta, green, and blue data are the results of the integral method according to~Ref.~\cite{Boyd:1996bx, Okamoto:1999hi, Borsanyi:2012ve}, respectively
}
\label{fig:quenched-result}
\end{center}
\end{figure}
The left panel of Fig.~\ref{fig:quenched-result} is our results for the
dimensionless interaction measure ($\Delta/T^4=(\varepsilon-3P)/T^4$) and the
dimensionless entropy density ($s/T^3=(\varepsilon+P)/T^4$) at~$T=1.65T_c$ as a
function of the dimensionless flow parameter~$\sqrt{8t}T$. The bold bars denote
the statistical errors, while the thin (light color) bars show the statistical
and systematic errors including the uncertainty
of~$\Lambda_{\overline{\text{MS}}}$. 
In the small $t$ region, the statistical error is
dominant  for both $\Delta/T^4$ and~$s/T^3$, while in the large $t$ region
the systematic error from~$\Lambda_{\overline{\text{MS}}}$ becomes significant
for~$s/T^3$.
 For instance, the statistical (systematic) errors of the data for $N_\tau=8$ are
 $2.5\%$ ($0.11\%$) for $\Delta/T^4$
and $0.83\%$ ($4.4\%$) for $s/T^3$ at $\sqrt{8t}T=0.40$.

The fiducial window discussed in~\textbf{Step 2} is indicated by the dashed
lines in~left panel of~Fig.~\ref{fig:quenched-result}. The lower limit, beyond
which the lattice discretization error grows, is set to
be~$\sqrt{8t_{\rm min}}=2a$, 
where we consider the size~$2a$ of our clover
leaf operator. The upper limit, beyond
which the smearing by the gradient flow exceeds the temporal lattice size, is
set to be $\sqrt{8t_{\rm max}}=1/(2T)=N_\tau a/2$.

Finally, we plot, in the right panel of Fig.~\ref{fig:quenched-result}, $\Delta/T^4$ and~$s/T^3$ after taking the continuum limit using the linear fit of the $N_\tau=6$, $8$,
and~$10$ data for $T/T_c=1.65$, $1.24$, and~$0.99$. 
For comparison,
the results of~Ref.~\cite{Boyd:1996bx, Okamoto:1999hi, Borsanyi:2012ve} obtained by the integral method
are shown by the magenta, green, and blue data in the right panel of~Fig.~\ref{fig:quenched-result}.
The results of the two different approaches are consistent
with each other within the statistical error.

\section{(2+1)-flavor QCD}
\subsection{strategy}
Now, let us extend the previous method to the full QCD system.
One property, which is no necessity of wave function renormalization factor of composite operators, is lost for the dynamical fermion system.
We have to take care of the renormalization for fermion fields.
Except for this and several trivial technical steps, the previous small flow-time expansion can straightforwardly apply to the full QCD system.  

Firstly, the fermion flow equation in terms of the Wilson flow-time must be solved in the case of full QCD.
The equation is given by
\beq
\partial_t \chi^f(t,x) &=& \Delta \chi^f(t,x), ~~~\chi^f(t=0,x)=\psi^f(x),\nonumber\\
\partial_t \bar{\chi}^f(t,x) &=& \bar{\chi}^f (t,x) \overset{\leftarrow}{\Delta}, ~~~\bar{\chi}^f (t=0,x)=\bar{\psi}^f(t,x),
\eeq
where $f$ denotes a label of the quark flavor and $\Delta \chi^f (t,a)=D_\mu D_\mu \chi^f(t,x)$.
Note that the covariant derivative refers to the flowed gauge field at the flow time $t$.

To obtain the thermal bulk quantity, which is independent of the imaginary time, introducing the random source fields and solving the adjoint flow equation proposed in Ref.~\cite{Luscher:2013cpa} are useful.

The EMT for the full QCD in the small flow-time limit is given by five dimension-$4$ operators in Ref.~\cite{Makino:2014taa} as follows:
\beq
T_{\mu\nu}^R(x) &=& \lim_{t\to0}\left\{ c_1(t)  \left[  O_{1 \mu \nu}(t,x)  - \frac{1}{4} O_{2 \mu \nu}(t,x)   \right]  \right. \nonumber\\
&&+c_2 (t) \left[ O_{2 \mu \nu}(t,x)  - \langle O_{2 \mu \nu}(t,x)  \rangle \right] \nonumber \\
&& +c_3 (t) \sum_{f=u,d,s} \left[ O^f_{3 \mu \nu}(t,x)  - 2 O^f_{4 \mu \nu}(t,x)  - \langle O^f_{3 \mu \nu}(t,x)  -2 O^f_{4 \mu \nu}(t,x)  \rangle  \right]\nonumber\\
&& +c_4 (t) \sum_{f=u,d,s} \left[ O^f_{4 \mu \nu}(t,x)  -\langle O^f_{4 \mu \nu}(t,x) \rangle  \right] \nonumber \\
&&+ c^f_5 (t) \left. \sum_{f=u,d,s} \left[  O^f_{5 \mu \nu}(t,x) - \langle O^f_{5 \mu \nu}(t,x) \rangle \right]     \right\} .
\eeq
The scalar coefficients $c_1 (t)$--$c^f_5 (t)$ within $1$loop order are given by the running coupling constant and running masses for each flavor and explicitly shown in Eqs.(4.60)--(4.64) in Ref.~\cite{Makino:2014taa}.
Note that the $c^f_5(t)$, which is related with the quark mass term, depends on the flavor.
The explicit form of each dimension-$4$ operator are given by
\beq
O_{1 \mu \nu} (t,x)&\equiv& G_{\mu \rho}^a (t,x) G_{\rho \nu}^a (t,x), \nonumber\\
O_{2 \mu \nu} (t,x)&\equiv& \delta_{\mu \nu} G_{\rho \sigma}^a (t,x) G_{\rho \sigma}^a (t,x), \nonumber\\ 
O^f_{3 \mu \nu} (t,x)&\equiv& \varphi^f (t) \bar{\chi}^f (t,x) \left( \gamma_\nu  \overset{\leftrightarrow}{D}_\nu +\gamma_\nu \overset{\leftrightarrow}{D}_\mu \right) \chi^f(t,x) ,\nonumber\\
O^f_{4 \mu \nu} (t,x)&\equiv&  \varphi^f (t) \delta_{\mu \nu} \bar{\chi}^f (t,x) \gamma_\rho  \overset{\leftrightarrow}{D}_\rho \chi^f (t,x), \nonumber\\
O^f_{5 \mu \nu} (t,x)&\equiv&  \varphi^f (t) \delta_{\mu \nu}  \bar{\chi}^f (t,x)   \chi^f (t,x),
\eeq
where $\overset{\leftrightarrow}{D}_\mu \equiv D_\mu - \overset{\leftarrow}{D}_\mu$.
The coefficient $\varphi(t)$ is introduced to cancel the renormalization factor for the fermion and defined by
\beq
\varphi^f (t) \equiv \frac{-6}{(4\pi)^2 t^2 \langle \bar{\chi}^f (t,x) \overset{\leftrightarrow}{D}_\mu \gamma_\mu \chi^f (t,x) \rangle} .\label{eq:varphi}
\eeq
This is numerically calculated by the lattice configurations at zero temperature simulation.

In full QCD case, three dimension-$4$ operators, $O_3$ -- $O_5$, are added.
To obtain the thermal quantities from the EMT, we additionally compute these expectation values.
These values can be summarized two types of expectation value as follows; 
\beq
t^f_{\mu \nu}(t) &\equiv& \frac{1}{N_\Gamma} \sum_x \langle \bar{\chi}^f (t,x) \gamma_\mu \left( D_\nu - \overset{\leftarrow}{D}_\nu  \right) \chi^f (t,x) \rangle,\label{eq:t-tensor} \\
s^f(t) &\equiv& \frac{1}{N_\Gamma} \sum_x \langle \bar{\chi}^f (t,x) \chi^f (t,x) \rangle,\label{eq:s-tensor}
\eeq
where $N_\Gamma$ is the lattice volume in lattice unit.

Now, let us summarize the procedure to calculate the EMT on the lattice for $(2+1)$-flavor QCD:\\
\textbf{Step 1:} Generate gauge configurations at~$t=0$ on a space-time lattice
with the lattice spacing~$a$ and the lattice size~$N_s^3\times N_\tau$ {\it{with dynamical fermions}}\\
\textbf{Step 2:} Solve {\it{both the gradient flow for the link variable and the adjoint fermion flow}} to obtain the
flowed field.\\
\textbf{Step 3-1:} Construct $U_{\mu\nu}(t,x)$ and~$E(t,x)$
in~Eqs.~(\ref{eq:(2)}) and~(\ref{eq:(3)}) in terms of the flowed link variables
and average over the gauge configurations at each~$t$.
\\
\textbf{Step 3-2:} Calculate $t_{\mu\nu}(t,x)$, $s(t,x)$ and $\varphi(t)$ in~Eqs.~(\ref{eq:t-tensor}), (\ref{eq:s-tensor}) and (\ref{eq:varphi}) using the flowed quark fields
and average over the gauge configurations at each~$t$, respectively.
\\
\textbf{Step 4:} 
Carry out an extrapolation toward $(a,t)=(0,0)$, first $a\to0$ and then $t\to0$ within the fiducial window of the flow time.

\subsection{Simulation details and preliminary results}
We performed QCD with $(2+1)$-flavor of quarks.
The Iwasaki gauge action and the standard $O(a)$-improved Wilson fermion are used in the simulation.
The lattice parameters are determined based on Ref.~\cite{Umeda:2012er}.
The hopping parameter is tuned to realize $m_{PS}/m_V=0.6337$ for $u,d$ quarks and $m_{PS}/m_V=0.7377$ for $s$ quark.
The lattice extent is $N_s^3 \times N_{\tau}=32^3 \times 8$ and the lattice bare gauge coupling and $c_{SW}$ are $\beta=1.973$ and $c_{SW}=1.669$.
The temperature is estimated to be $280$MeV.

The number of configuration is still $11$, where we take $100$ Montecarlo trajectories between measurements.
The error bar denotes the standard statistical error.

\begin{figure}[h]
\begin{center}
\includegraphics[scale=0.5]{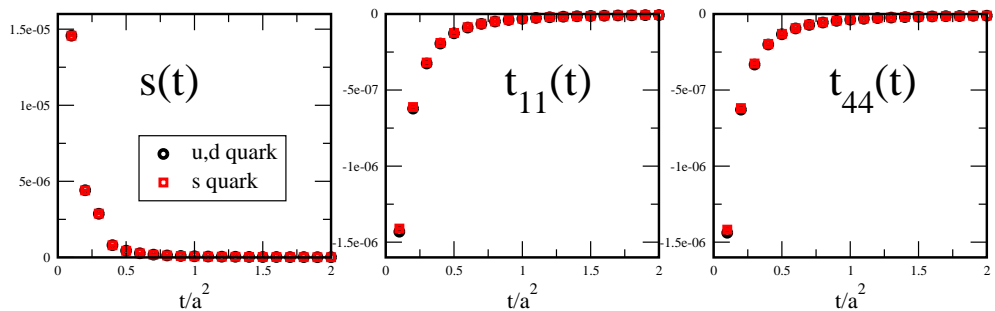}
\caption{Flow time dependence of the $s(t)$(left), $t_{11}$(middle) and $t_{44}$(right) with $\beta=1.973$, $c_{SW}=1.669$, $\kappa_{u,d}=0.1361$, $\kappa_s=0.1354$ on $32^3 \times 8$ lattices. Circle (black) and square (red) symbols denote $u,d$ and $s$ quarks on each panel, respectively.
}
\label{fig:fermi-flow}
\end{center}
\end{figure}

Figure~\ref{fig:fermi-flow} shows the preliminary results for $t_{\mu \nu}(t)$ and $s(t)$ in Eqs.~(\ref{eq:t-tensor}) and (\ref{eq:s-tensor}).
The circle (black) and square (red) symbol present $u,d$ and $s$ quarks on each panel, respectively.
Here, $s(t)$ is the chiral condensate. It decreases to almost zero as expected, since the present temperature is higher than $T_c$.

The statistical error is a few \% order in these operators with only $11$ configurations. 
We consider that the method looks promising to obtain the thermal quantities even in the full QCD simulation, although the careful estimation of the autocorrelation should be done.
Further simulations will give a precise determination of them in near future.

\section*{Acknowedgement}
We would like to thank M.~Asakawa, T.~Hatsuda, T.~Iritani and M.~Kitazawa for useful discussions. We are also grateful to K.~Kanaya and H.~Matsufuru for their helps of the code development.

\end{document}